\documentclass[]{aa}
\usepackage{txfonts}
\usepackage{natbib}
\usepackage{graphicx}
\bibpunct{(}{)}{;}{a}{}{,} 

\newcommand{\ind}[1]{_{\mathrm{#1}}}
\newcommand{\diff}{\mathrm{d}}
\def\numax{\nu\ind{max}}
\def\dnuenv{\delta\nu\ind{env}}
\def\Dnu{\Delta\nu}
\def\dnumoy{\langle\Delta\nu\rangle}

\newcommand{\BV}{Brunt-V\"ais\"al\"a}
\newcommand{\NBV}{N\ind{BV}}
\newcommand{\spacing}{\delta\nu_{\mathrm{g,1}}}
\renewcommand{\ng}{n\ind{g}}
\newcommand{\tgun}{\Delta T_1}
\newcommand{\tgobs}{\Delta T\ind{obs}}
\newcommand{\tgdeux}{\Delta T_2}
\newcommand{\per}{T}
\newcommand\Kepler{\emph{Kepler}}

\begin{document}
\title{Mixed modes in red-giant stars observed with CoRoT
\thanks{The CoRoT space mission, launched on 2006 December 27, was developed
and is operated by the CNES, with participation of the Science
Programs of ESA, ESA's RSSD, Austria, Belgium, Brazil, Germany and
Spain.}}

\titlerunning{Mixed modes in red-giant stars}
\author{B. Mosser\inst{1}
\and C. Barban\inst{1} \and J. Montalb\'an\inst{2} \and P.G.
Beck\inst{3} \and A. Miglio\inst{2,4} \and K. Belkacem\inst{5,1}
\and M.J. Goupil\inst{1} \and S. Hekker\inst{4,6}\and J. De
Ridder\inst{3} \and M.A Dupret\inst{2} \and Y. Elsworth\inst{4}
\and A. Noels\inst{2} \and F. Baudin\inst{5} \and E.
Michel\inst{1} \and R. Samadi\inst{1} \and M. Auvergne\inst{1}\and
A. Baglin\inst{1}\and C. Catala\inst{1}}

\offprints{B. Mosser}

\institute{LESIA, CNRS, Universit\'e Pierre et Marie Curie, Universit\'e Denis Diderot,
Observatoire de Paris, 92195 Meudon cedex, France; \email{benoit.mosser@obspm.fr}
\and
Institut d'Astrophysique et de G\'eophysique, Universit\'e de Li\`ege, All\'ee du 6 Ao\^ut, 17 B-4000 Li\`ege, Belgium
\and
Instituut voor Sterrenkunde, K. U. Leuven, Celestijnenlaan 200D, 3001 Leuven, Belgium
\and
School of Physics and Astronomy, University of Birmingham, Edgbaston, Birmingham B15 2TT, United Kingdom
\and
Institut d'Astrophysique Spatiale, UMR 8617, Universit\'e Paris XI, B\^atiment 121, 91405 Orsay Cedex, France
\and
Astronomical Institute `Anton Pannekoek', University of Amsterdam, Science Park 904,
1098 XH Amsterdam, The Netherlands
}

\abstract{The CoRoT mission has provided thousands of red-giant
light curves. The analysis of their solar-like oscillations allows
us to characterize their stellar properties.}
{Up to now, the global seismic parameters of the pressure modes
remain unable to distinguish red-clump giants from members of the
red-giant branch. As recently done with Kepler red giants, we
intend to analyze and use the so-called mixed modes to determine
the evolutionary status of the red giants observed with CoRoT. We
also aim at deriving different seismic characteristics depending
on evolution.}
{The complete identification of the pressure eigenmodes provided
by the red-giant universal oscillation pattern allows us to aim at
the mixed modes surrounding the $\ell$=1 expected
eigenfrequencies. A dedicated method based on  the envelope
autocorrelation function is proposed to analyze their period
separation.
}%
{We have identified the mixed-mode signature separation thanks to
their pattern compatible with the asymptotic law of gravity modes.
We have shown that, independent of any modelling, the g-mode
spacings help to distinguish the evolutionary status of a
red-giant star. We then report different seismic and fundamental
properties of the stars, depending on their evolutionary status.
In particular, we show that high-mass stars of the secondary clump
present very specific seismic properties. We emphasize that stars
belonging to the clump were affected by significant mass loss. We
also note significant population and/or evolution differences in
the different fields observed by CoRoT.
}%
{}


\maketitle

\voffset = 1.5cm
\section{Introduction\label{introduction}}

The CoRoT and \Kepler\ missions have revealed solar-like
oscillation in thousands of red-giant stars. This gives us the
opportunity to test this important phase of stellar evolution, and
provides new information in stellar and galactic physics
\citep{2009A&A...503L..21M,2011Natur.471..608B}. Thanks to the
dramatic increase of information recently made available
\citep{2009Natur.459..398D,2009A&A...506..465H,2010A&A...517A..22M,2010A&A...522A...1K,
2010ApJ...723.1607H}, we have now a precise view of pressure modes
(p modes) corresponding to oscillations propagating essentially in
the large convective envelopes. Gravity modes (g modes) may exist
in all stars with radiative regions. They result from the trapping
of gravity waves, with buoyancy as a restoring force. In red
giants, gravity waves propagating in the core have high enough
frequencies to be coupled to pressure waves propagating in the
envelope \citep{2009A&A...506...57D}. The trapping of such waves
with mixed pressure and gravity character gives the so-called
mixed modes. The coupling insures non-negligible oscillation
amplitudes in the stellar photosphere, hence the possible
detection of these mixed modes.

Observationally, mixed modes were first identified in white-dwarf
oscillation spectra \citep{1991ApJ...378..326W}. They have also
been observed in sub-giant stars, first with ground-based
observations
\citep{2005A&A...434.1085C,2007ApJ...663.1315B,2010ApJ...713..935B},
then recently in \Kepler\ and CoRoT fields
\citep{2010ApJ...713L.169C,2010A&A...515A..87D}. In red giants,
they were first suspected by \cite{2010ApJ...713L.176B}. Their
presence significantly complicates the fit of the p modes observed
in CoRoT giants \citep{2010A&A...520A..60H,2010AN....331.1016B}
and they were identified as outliers to the universal red-giant
oscillation spectrum \citep{2011A&A...525L...9M}. Recent modelling
of a giant star observed by the \Kepler\ mission had to take their
presence into account \citep{2011arXiv1105.1076D}. Finally, they
have been firmly identified in \Kepler\ data
\citep{2011Sci...332..205B}, making the difference clear between
giants burning hydrogen in shell or helium in the core
\citep{2011Natur.471..608B}.

The theoretical analysis of these mixed modes in red giants has
been performed prior to observations
\citep{2001MNRAS.328..601D,2004SoPh..220..137C,2009A&A...506...57D}.
Using a non-radial non-adiabatic pulsation code including a
non-local time-dependent treatment of convection and a stochastic
excitation model, \cite{2009A&A...506...57D} have computed the
eigenfrequencies and the mode heights in several red-giant models.
They have shown that mixed modes have much larger mode inertias
than p modes, hence present longer lifetimes and smaller
linewidths. They were able to identify different regimes,
depending on the location of the models on the red-giant branch.
Recently, \cite{2010ApJ...721L.182M} proposed to use the
oscillation spectrum of dipole modes to discriminate between
red-giant branch (RGB) and central-He burning (clump) evolutionary
phase of red giants. This illustrates the fact that observing
p-g-mixed modes and identifying their properties give us a unique
opportunity to analyze the cores of red giants, since the g
component is highly sensitive to the core condition
\citep{2009A&A...506...57D}.

In this paper, we present and validate an alternative method to
\cite{2011Natur.471..608B} to detect and identify mixed modes in
red giants. We use it to analyze red-giant stars in two different
fields observed by CoRoT and show that they present different
properties: mixed modes do not only allow us to distinguish
different evolutionary status, they can also show different
population characteristics. We also assess clear observational
differences between the fundamental and seismic properties of the
red-giant stars, depending on their evolutionary status. We show
for instance that the observation of mixed modes opens a new way
to study the mass loss at the tip of the RGB. We have also clear
indication that mixed modes in red giants are sensitive to
chemical composition gradients in the deep interior, as they are
in SPB and {$\gamma$} Doradus stars \citep{2008MNRAS.386.1487M}.
Their study will definitely boost both stellar and galactic
physics.

\begin{figure*}
\includegraphics[width=17.5cm]{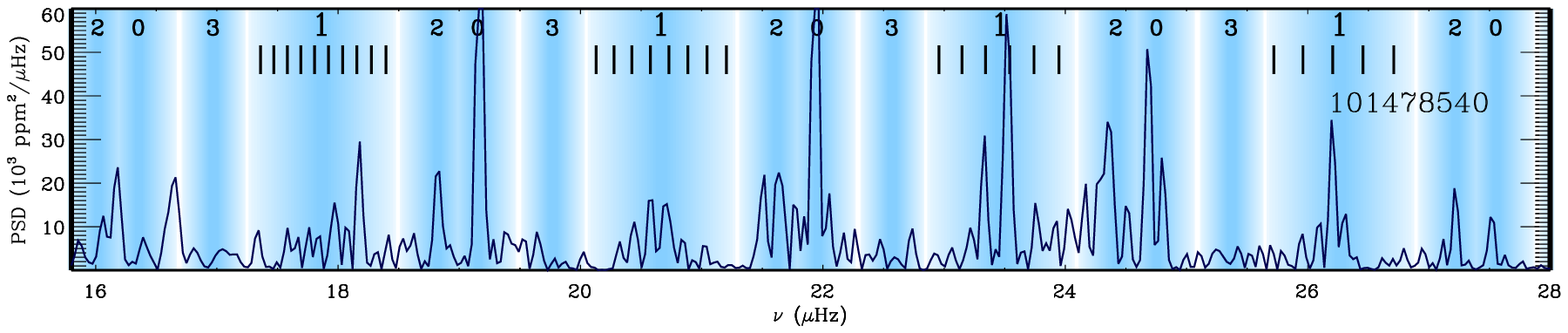}
\vskip-0.15cm
\includegraphics[width=17.5cm]{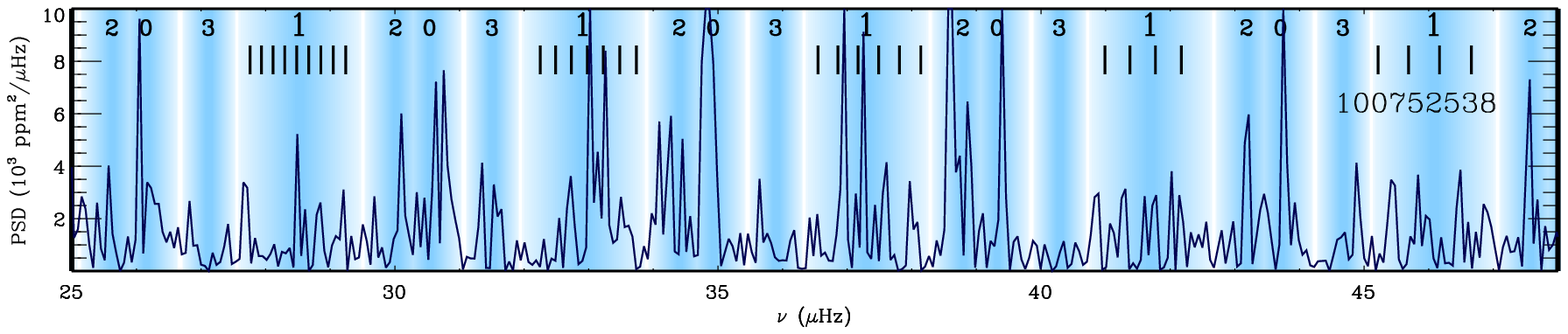}
\vskip-0.15cm
\includegraphics[width=17.5cm]{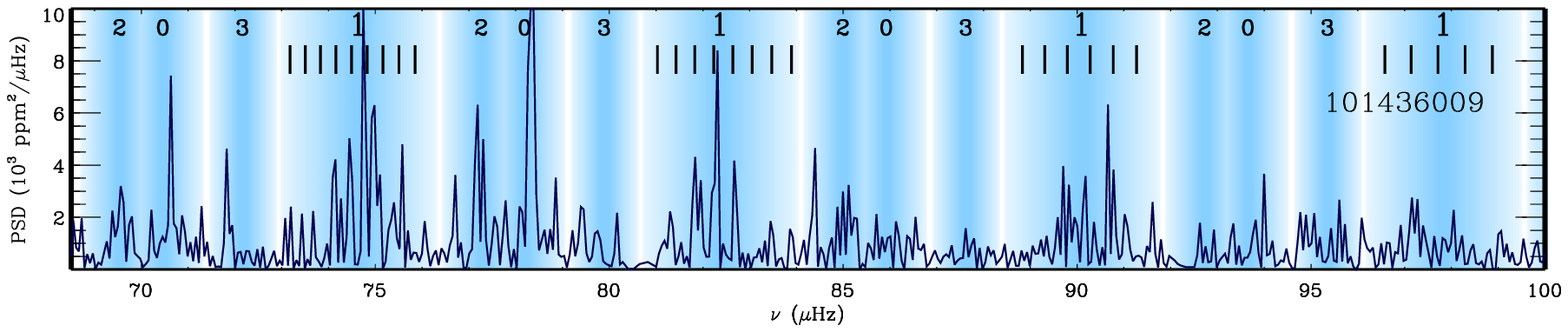}
\caption{Shaded regions of these \emph{bar code spectra} reproduce
the universal red-giant oscillation pattern and identify the
location of the different harmonic degrees for three CoRoT targets
(ID given on the right side); the more dense the background, the
higher the probability to have the short-lived p mode realized
there. Black dashes in the $\ell=1$ ridges indicate the asymptotic
spacing of the mixed modes as derived from Eq.~\ref{tassoul_un};
they do not give their exact eigenfrequencies, since the
asymptotic relation does not account for the mode bumping of mixed
modes.
\label{fig-fit}}
\end{figure*}

\section{Analysis\label{validation}}

\subsection{Mixed-mode pattern}

According to \cite{2009A&A...506...57D}, each $\ell=1$ ridge is
composed of a pressure mode surrounded by mixed modes with a
pattern mostly dominated by the g-mode component. According to the
asymptotic description \citep{1980ApJS...43..469T}, both p modes
and g modes present well organized patterns. For p modes, the
organization is a comb-like structure in frequency, for g modes,
regularity is observed in period. Mixed modes in giants are
supposed to exhibit the well-organized pattern of g modes. In
order to detect them in an automated way, we have searched for
this asymptotic pattern:
\begin{equation}
\per_{\ng, \ell} = %
{2\pi^2 (\ng + \alpha_\ell) \over \sqrt{\ell (\ell +1)}} \ \left[
\int_{\mathrm{core}} \!\!{{\NBV \over r}\; \diff r } \right]^{-1}
= %
{(\ng + \alpha_\ell)\  \Delta T\ind{g} \over \sqrt{\ell (\ell
+1)}} \label{tassoulg}
\end{equation}
with $\ng$ the gravity radial order, $\alpha_\ell$ an unknown
constant and $\NBV$ the \BV\ frequency. The period spacing $\Delta
T\ind{g}$ is the equivalent for g modes of the large separation
for p modes.

\subsection{Periodic spacings}

We have first restricted our attention to $\ell=1$ modes since
mixed modes with $\ell=2$ are not supposed to be as clearly
visible \citep{2009A&A...506...57D}. Furthermore, the close
proximity of $\ell=2$ modes to radial modes complicates the
analysis. For the $\ell=1$ mixed modes, the periodic pattern
simply becomes:
\begin{equation}
\per_{\ng, 1} = \left[ \ng + \alpha_1 \right] \tgun , \hbox{ with
\  } \tgun=  \Delta T\ind{g} / \sqrt{2} . \label{tassoul_un}
\end{equation}
The period spacing $\tgun$, linked to the integral of the \BV\
frequency by Eq.~\ref{tassoulg}, can be deduced from the frequency
spacing in the Fourier spectrum. To measure this spacing, we need
to focus on the mixed modes. We are able to do this using the
method introduced in \cite{2011A&A...525L...9M} which allows for a
complete and automated mode identification. We then use the
envelope autocorrelation function \citep{2009A&A...508..877M} to
derive the period spacing. The method is described in the
Appendix. As for the method presented by
\cite{2011Natur.471..608B}, it derives a period $\tgobs$ less than
$\tgun$, due to the bumping of mixed modes. We have estimated
that, for the 5-month long CoRoT time series, we measure $\tgobs
\simeq \tgun / 1.15$ (see Appendix).

\begin{table*}[ht]
\caption{Typical parameters of mixed modes\label{param}}
\begin{tabular}{ccccccccc}
\hline
$\dnumoy$&$\numax$&$\dnuenv$&$\tgobs$&$\tgun$ proxy&$\ng; \Delta\ng$&$\ng; \Delta\ng$ &$\ng; \Delta\ng$    & $\mathcal T$\\
         &        &         &        &             &at $\numax-\dnuenv/2$&at $\numax$&at $\numax+\dnuenv/2$ & \\
$\mu$Hz  & $\mu$Hz&$\mu$Hz  &   s    &   s         &  & & & day \\
\hline
\multicolumn{9}{c}{Red-clump stars}\\
  4.0  & 32 & 13 & 250  & 287  &137; 5  &108; 3  & 89; 2 & 78 \\
  7.0  & 73 & 28 & 150  & 173  & 97; 3  & 78; 2  & 66; 1 & 25 \\
\hline
\multicolumn{9}{c}{Red-giant branch stars}\\
  2.5  & 19 &  8 & 800  & 920  &  73; 3  & 57; 2  & 46; 1 & 70 \\
  4.0  & 32 & 13 &  80  &  92  & 429;17  &339;10  &281; 7 & 245\\
  6.0  & 60 & 23 &  60  &  69  & 300; 9  &241; 6  &201; 4 & 93 \\
  9.0  &102 & 38 &  60  &  69  & 174; 5  &142; 3  &119; 2 & 32 \\
\hline
\end{tabular}

The frequency $\numax$ corresponds to the maximum oscillation
signal; $\dnuenv$ indicates the full width at half-maximum of the
observed excess oscillation power \citep{2010A&A...517A..22M}. The
proxies of the period spacing $\tgun$ are estimated as $1.15\,
\tgobs$.  The values of $\ng$ are derived from
Eq.~\ref{tassoul_un}, with the parameter $\alpha_{\ell=1}=0$.
$\Delta\ng$ corresponds to the maximum number of observable mixed
modes in a $0.25\,\Dnu$-wide interval around each pure p mode.
$\mathcal T$ indicates the observational length required to
measure the g-mode spacing according to the Shannon criterion.
\end{table*}

\subsection{Positive detection}

The analysis has been performed on CoRoT red giants previously
analyzed in \cite{2009A&A...506..465H} and
\cite{2010A&A...517A..22M}. These stars were observed continuously
during 5 months in 2 different fields respectively centered on the
Galactic coordinates (37$^\circ$, $-07^\circ$45\arcmin ) and
(212$^\circ$, $-01^\circ$45\arcmin). Targets with a too low
signal-to-noise ratio were excluded (see the Appendix). We
therefore only considered stars with accurate global seismic
parameters and labelled as the $\mathcal{N}_3$ set in
\cite{2010A&A...517A..22M}.

\begin{figure}
\includegraphics[width=8.9cm]{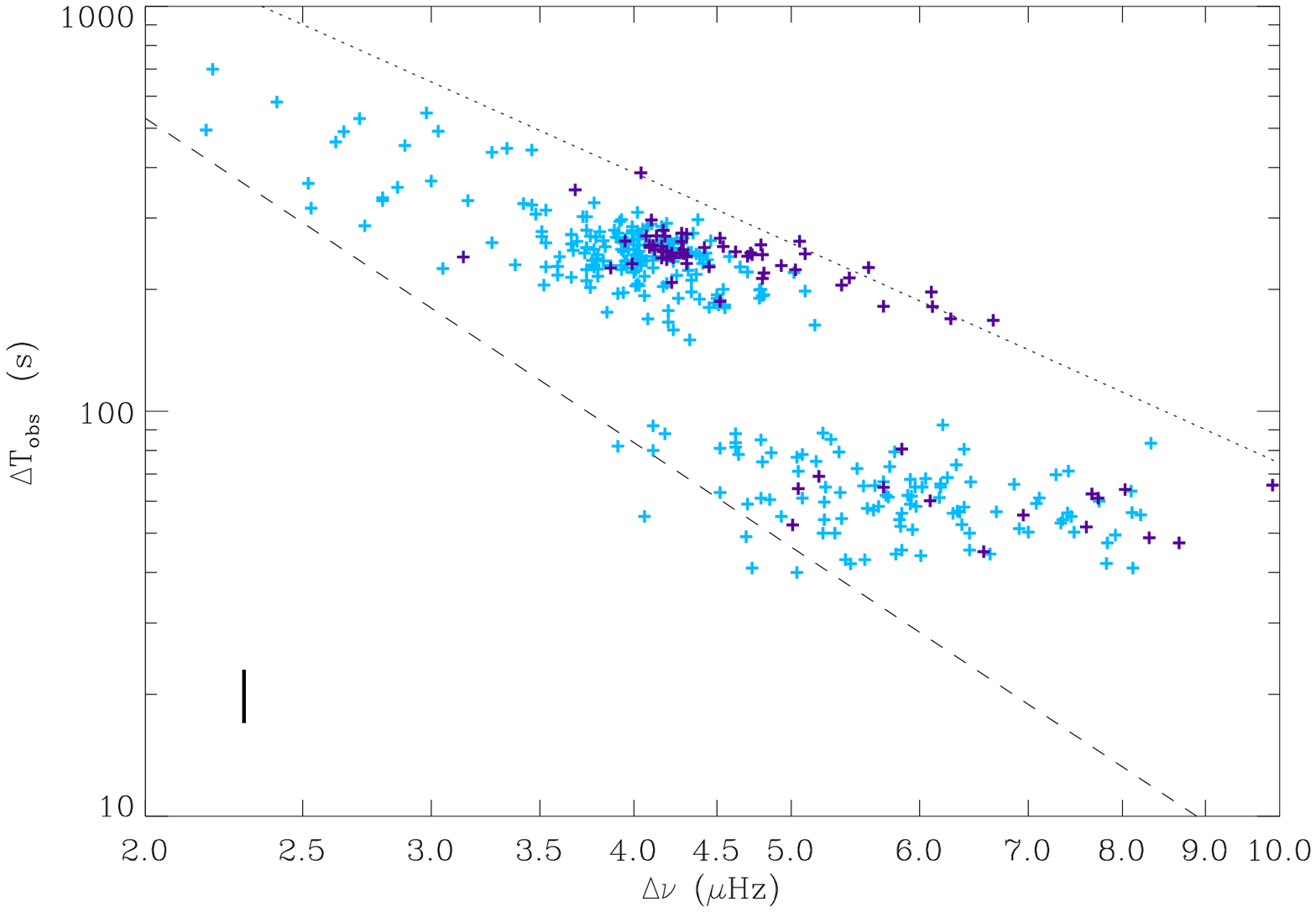}
\caption{Period separation of mixed modes $\tgobs$ as a function
of the mean frequency separation of the pressure modes $\dnumoy$.
Blue (purple) crosses represent the targets observed in the center
(anticenter) direction. The dashed line indicates the frontier
below which the frequency resolution is not fine enough for
deriving $\tgobs$ and the dotted line represents the location of
the spurious signature that would correspond to the small
separation $\delta\nu_{02}$ \citep{2011A&A...525L...9M}. The two
separate domains correspond to the location of $\tgun$ expected by
\cite{josefina} and observed by \cite{2011Natur.471..608B}. The
vertical thick line indicates the mean 1-$\sigma$ error bars (the
error bar on the large separation $\Dnu$ is in fact very small).
\label{scalingT}}
\end{figure}

Thanks to the method exposed in the Appendix, we have analyzed all
red-giant spectra in an automated way. We have then checked all
results individually. This allowed us to discard a few false
positive results, and then to verify that the asymptotic expansion
of the g modes (Eq.~\ref{tassoul_un}) gives an accurate
description of the mixed modes since it is able to reproduce their
spacings. The irregularities of the spacings are discussed in
Section \ref{period_spacing} and in the Appendix. A few examples
are given in Fig.~\ref{fig-fit}: we have overplotted on red-giant
oscillation spectra the expected location of the p-mode pattern
derived from \cite{2011A&A...525L...9M} and we have indicated the
frequency separation of the mixed modes derived from the
asymptotic description. Because of the mode bumping due to the
coupling, the g-mode asymptotic expression is not able to derive
the exact location of the mixed-mode eigenfrequencies, but we see
that the spacings reproduce the observations. This agreement is
certainly related to the fact that, according to
Eq.~\ref{tassoul_un}, very high values of the gravity radial
orders are measured, typically above 60 and up to 400
(Table~\ref{param}). We were then able to provide a diagram
representing the period separation $\tgobs$ of the mixed modes as
a function of the large frequency separation $\Dnu$ of the
pressure modes (Fig.~\ref{scalingT}).

The 5-month long CoRoT time series provide a frequency resolution
of about 0.08\,$\mu$Hz, accurate enough for detecting the
mixed-modes in the red clump but limiting their possible detection
at lower frequency. Therefore, we have taken care of possible
artefacts. We have excluded the region of the $\dnumoy$ - $\tgobs$
diagram limited by the frequency resolution (Fig.~\ref{scalingT}).
We have also taken care of the possible confusion with the small
separation $\delta\nu_{02}$, since in given frequency ranges its
signature can mimic a g-mode spacing (dotted line in
Fig.~\ref{scalingT}). Due to the low visibility of $\ell=3$ modes,
it appeared useless to test the influence of their pattern
combined with radial or dipole modes.

The criterion of a positive detection of the mixed modes in at
least two frequency ranges allowed us to discriminate regular
period spacings from regular frequency spacings as caused by
rotation. We also examined the possible confusion with the ragged,
speckle-like, appearance of the structure due to the limited
lifetime of the modes. Simulations have shown that the threshold
level provided by the method \citep{2009A&A...508..877M} combined
with the detection in multiple frequency ranges excludes spurious
signature.

Finally, we have identified an $\ell=1$ mixed-modes pattern in
about 42\,\% of the targets (387 out of a total of 929), and for
more than 75\,\% of the stars with a R magnitude brighter than 13.
The few remaining bright stars do not present a reliable
signature, either because their spectrum is free of mixed modes,
or because the signature was not found reliable according to the
threshold detection level. When detected, the pattern follows
closely both the asymptotic expression of gravity modes
\citep{1980ApJS...43..469T} and the description made by
\cite{2009A&A...506...57D}. The peaks generating these spacings
are visible in a frequency range of about $0.25\,\Dnu$ around the
expected pure p-mode eigenfrequencies $\nu_{n,1}$. This represents
typically, at the peak of the red clump distribution
($\Dnu\simeq4\ \mu$Hz and $\numax \simeq 32\,\mu$Hz), from two to
ten mixed modes (Table~\ref{param}). Due to the frequency
dependence derived from Eq.~\ref{tassoul_un}, the number of
observable modes varies very rapidly, as observed by
\cite{2011Sci...332..205B} and made explicit in Table~\ref{param}.

\begin{figure}
\includegraphics[width=8.9cm]{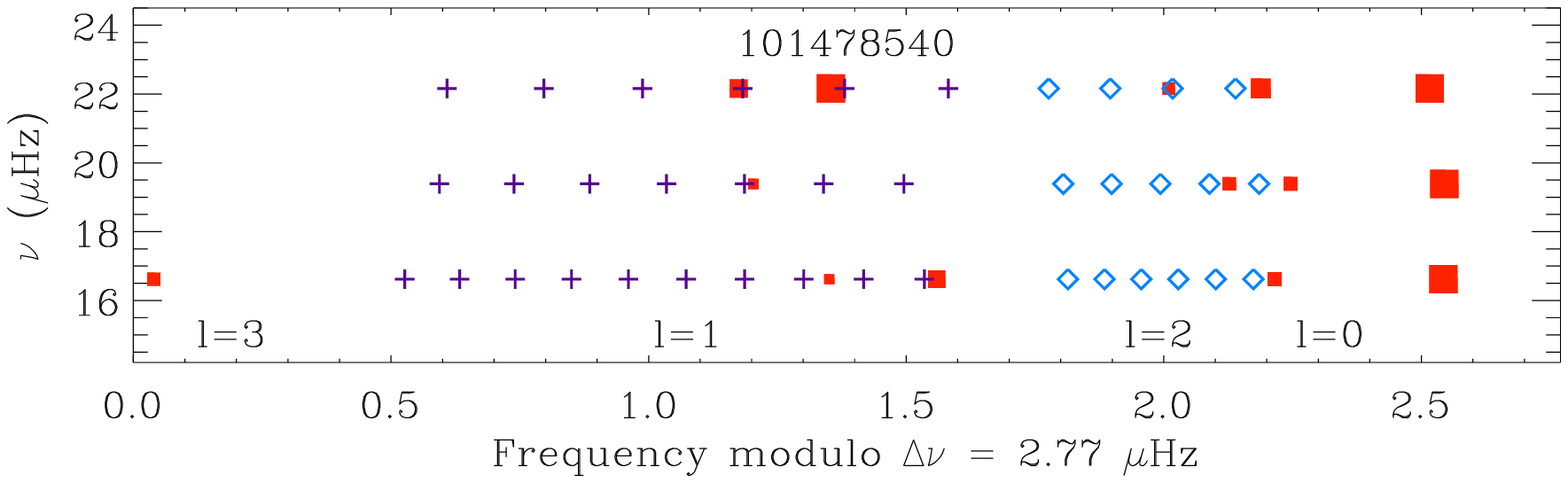}
\vskip 0.3cm
\includegraphics[width=8.9cm]{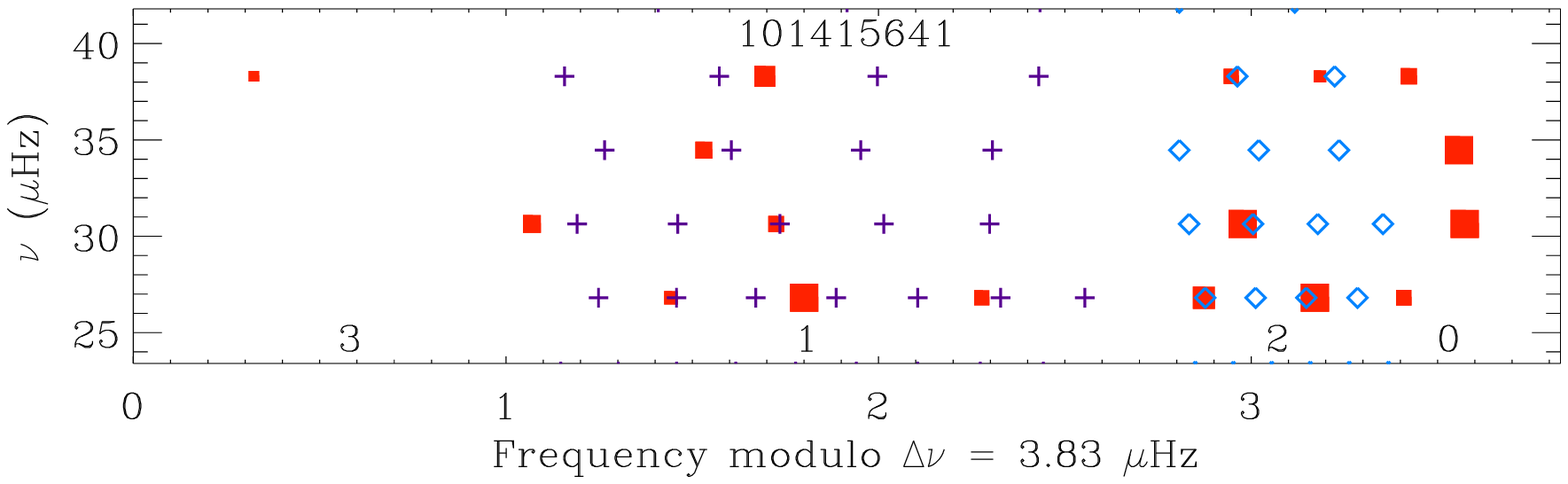}
\vskip 0.3cm
\includegraphics[width=8.9cm]{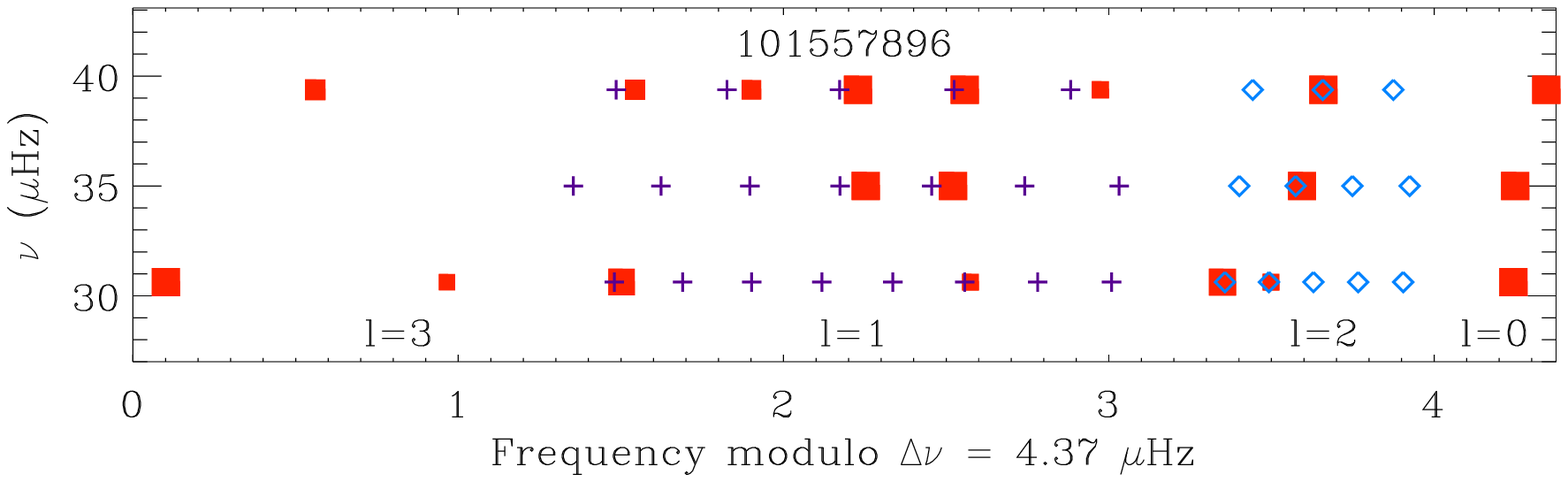}
\vskip 0.3cm
\includegraphics[width=8.9cm]{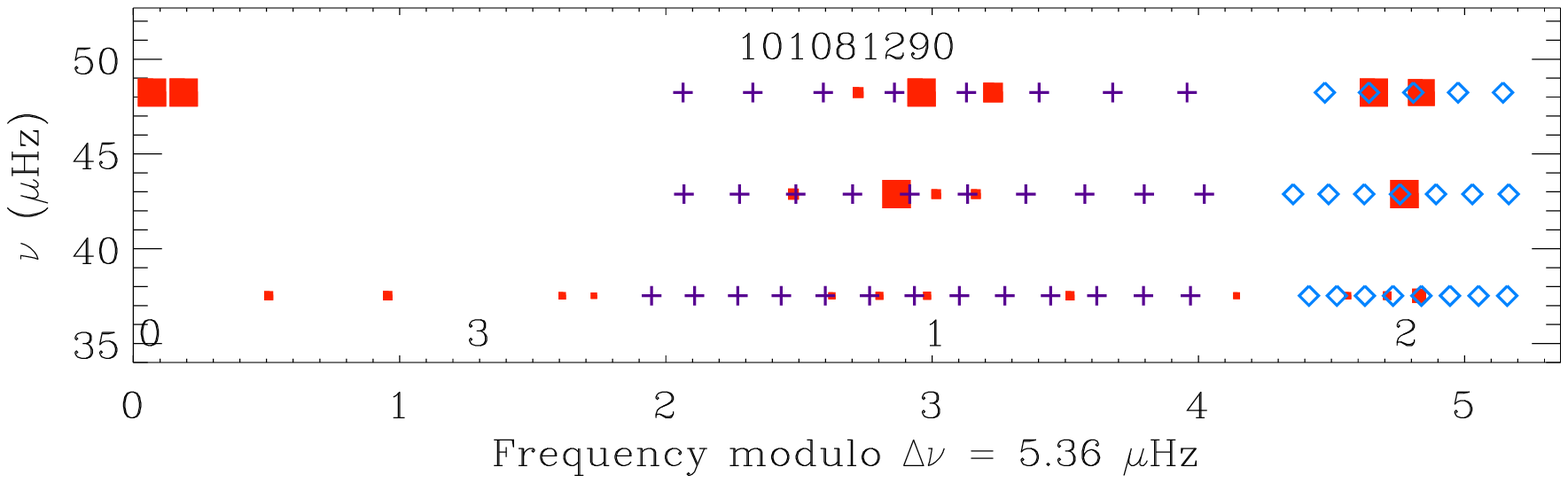}
\vskip 0.3cm
\includegraphics[width=8.9cm]{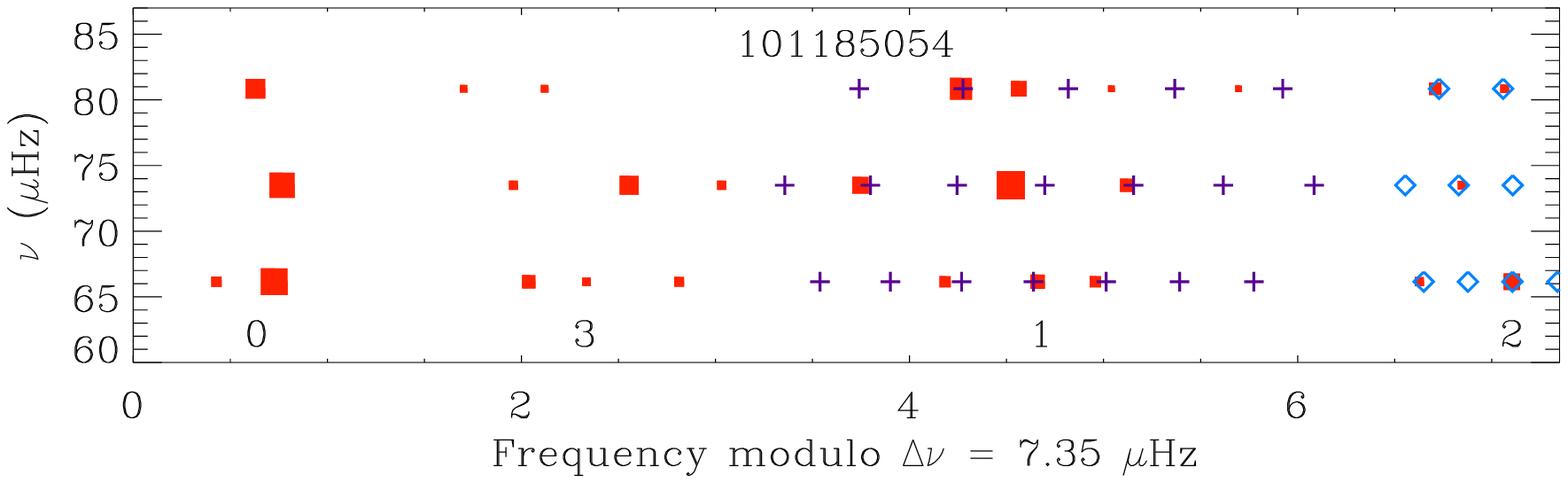}
\caption{\'Echelle diagrams around $\numax$ for different targets
sorted by increasing large separations. The red squares indicate
the observed peaks, selected with a height-to-background ratio
above 6. Dark blue crosses indicate the expected location of the
$\ell=1$ mixed modes strictly following the asymptotic expression;
light blue diamonds are for $\ell=2$. Similarly to
Fig.~\ref{fig-fit}, crosses and diamonds indicate the mean
\emph{spacing} of the mixed modes only, not their exact location.
 \label{mixed2}}
\end{figure}

According to \cite{2009A&A...506...57D}, $\ell=2$ mixed modes are
expected to be much more damped than $\ell=1$ modes. However,
their presence is clearly identified for about a third of the
targets (Fig.~\ref{mixed2}). The agreement of their mean
separation with $\tgdeux \simeq \tgun /\sqrt{3}$
(Eq.~\ref{tassoulg}) is clear enough to be disentangled from the
speckle-like aspect of short-lived modes. Their observation will
give strong constraints on the inner envelope, where g modes are
coupled to p modes, the coupling being highly sensitive to the
respective location of  the \BV\ and $\ell=2$ Lamb frequency. The
high density of mixed modes, due to the high values of the radial
gravity orders observed, makes it possible to achieve this
high-resolution analysis. We observe in most cases groups of only
two $\ell=2$ mixed modes, rarely three, spread in a frequency
range of about $0.04\,\Dnu$.
A further consequence of the observation of $\ell=2$ mixed modes
is a correction to the asymptotic laws reported for the small
separation $\delta\nu_{02}$ by \cite{2010ApJ...723.1607H} and
\cite{2011A&A...525L...9M}; they are correct, but only indicative
of the barycenter of the $\ell=2$ modes, if mixed.

\section{Discussion\label{results}}

\subsection{Period spacing\label{period_spacing}}

The measurement of mixed-mode spacings varying as $1/\nu^2$
(Eq.~\ref{spacing}) validates the use of the asymptotic law of g
modes, even if a detailed view of the mixed modes indicates
irregularities in their spacings (Fig.~\ref{mixed2}). These shifts
may be interpreted as a modulation due to the structure of the
core with a sharp density contrast compared to the envelope, as
observed in white dwarfs. The determination of $\tgun$ will allow
modelers to define the size of the radiative core region. The
direct measurement of the individual shifts $\tgun (\ng)$ will
give access to the core stratification
\citep{2008MNRAS.386.1487M}, as well as the observation (or not)
of $\ell=2$ mixed modes that test a different cavity.

The method developed in this paper presents many interesting
characteristics compared to the method used by
\cite{2011Natur.471..608B}. First, it is directly applicable in
the Fourier spectrum and does not require the power spectrum to be
expressed in period. The method is fully automated, since it is
coupled to the identification of the spectrum based on the
universal pattern, and it includes a systematic search of periodic
spacings that are not related to the p-mode pattern. Based on an
$H_0$ test, it intrinsically includes a statistical test of
reliability. This test defines a threshold level that makes the
method efficient even at low signal-to-noise ratio. Its most
interesting property certainly consists in its ability to derive
the measurement of the variation of $\tgun$ with frequency, as the
EACF method used for p modes \citep{2009A&A...508..877M,
2010AN....331..944M}. In fact, the method was developed in
parallel to the one mainly presented in
\cite{2011Natur.471..608B}, and gave similar results that
confirmed the detection of mixed modes in \Kepler\ giants.

\subsection{Red-clump and red-giant branch stars}

Red-clump stars have been characterized in previous work
\citep[e.g. Fig. 5 of][]{2010A&A...517A..22M}. They contribute to
a distribution in $\Dnu$ with a pronounced accumulation near
4\,$\mu$Hz (equivalent to the accumulation near $\numax\simeq
32\,\mu$Hz). However, population analysis made by
\cite{2009A&A...503L..21M} has shown that this population of stars
with $\Dnu \simeq 4\,\mu$Hz effectively dominated by red-clump
stars also contains a non-negligible fraction ($\simeq 30$\,\%) of
RGB stars. None of the global parameters of p mode (neither $\Dnu$
nor $\numax$) is able to discriminate between RGB and red-clump
stars. However, as shown by \cite{2011Natur.471..608B} with
\Kepler\ data, the examination of the relation between the p-mode
and g-mode spacings shows two regimes (Fig.~\ref{scalingT}). The
signature of the red-clump stars piling up around $\dnumoy\simeq
4\,\mu$Hz is visible with $\tgobs \simeq 250$\,s. Another group of
stars has $\tgobs$ values lower than 100\,s. Regardless of any
modelling, this gives a clear signature of the difference between
red-clump stars that burn helium in their core and stars of the
red-giant branch that burn hydrogen in a shell
\citep{2010ApJ...721L.182M}. The agreement with theoretical values
is promising \citep{josefina}.

The contribution of the red-clump stars in the $\Dnu$-$\tgobs$
diagram presented in Fig.~\ref{scalingT} is unambiguous. Stars
with large separation larger than the clump value (about
4\,$\mu$Hz) are located on the ascending red-giant branch or
members of the secondary clump
\citep{1999MNRAS.308..818G,2009A&A...503L..21M}. At lower
frequency than the clump, using mixed modes to disentangle the
evolutionary status is more difficult. According to the
identification of the clump provided by the distributions of the
large separation \citep{2010A&A...517A..22M}, we assume that the
giants with $\Dnu\le 3.5\,\mu$Hz and $\tgobs \ge 200$\,s belong to
the RGB. The following analysis shows further consistent
indications.

We finally note that the detection of RGB stars having a large
separation similar to the peak of the clump stars ($\simeq
4\,\mu$Hz) is only marginally possible, due to insufficient
frequency resolution. According to Table \ref{param}, more than
200 days are necessary to resolve the mixed modes, while CoRoT
runs are limited to about 150 days.

\subsection{Mode lifetimes and heights}

We have remarked that, as predicted by \cite{2009A&A...506...57D},
the lifetimes of the mixed modes trapped in the core are much
larger than the lifetimes of radial modes (Fig.~\ref{fig-fit}). In
most cases, the mixed modes are not resolved. When the mixed modes
are identified, they corresponds in most cases to a comb-like
pattern of thin peaks, without a larger and broader peak that
could correspond to the pure $\ell=1$ pressure mode. However,
since the detection of mixed modes is only achieved for a limited
set of stars, one cannot exclude that some red giants only show
pure $\ell=1$ pressure modes.

\begin{figure}
\includegraphics[width=8.9cm]{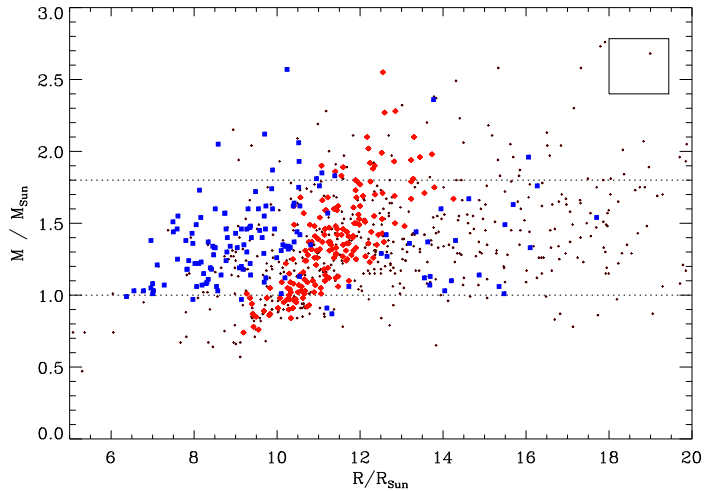}
\caption{Asteroseismic mass as a function of the asteroseismic
radius, with indication of the evolutionary status derived from
the mixed-mode spacing: blue squares for RGB stars, red diamonds
for clump stars; stars without clear measurement of $\tgobs$ are
marked with a small cross. The rectangle indicates the mean
1-$\sigma$ error bars. The dotted lines at respectively 1 and
1.8$\, M_\odot$ correspond to the limits defined in the text.
 \label{MR}}
\end{figure}

\begin{figure}
\includegraphics[width=8.9cm]{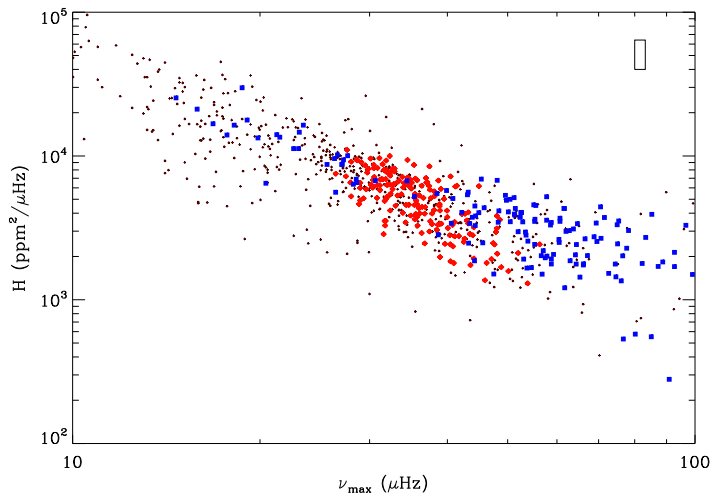}
\caption{Mean mode height as a function of $\numax$. Same color
code as in Fig.~\ref{MR}. The rectangle indicates the mean
1-$\sigma$ error bars.
 \label{height}}
\end{figure}

\begin{figure}
\includegraphics[width=8.9cm]{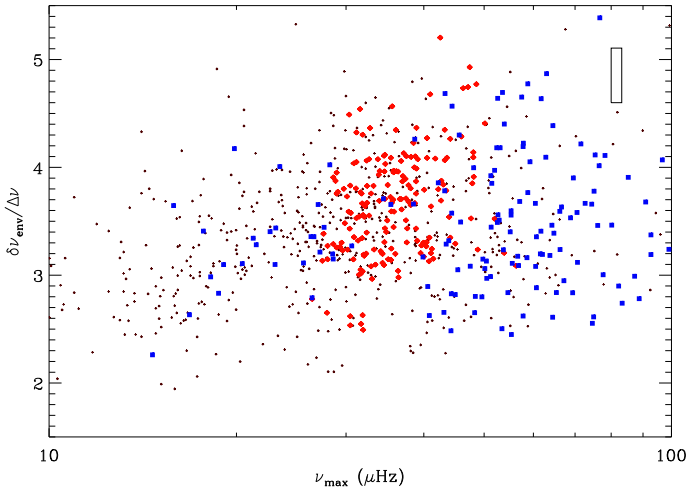}
\caption{Proxy of the number of observable modes, given by the
ratio $\dnuenv / \Dnu$, as a function of $\numax$. Same color code
as in Fig.~\ref{MR}. The rectangle indicates the mean 1-$\sigma$
error bars.
 \label{envel}}
\end{figure}

\subsection{Mass-radius relation}

The possibility to distinguish the evolutionary status allows us
to refine the ensemble asteroseismic analysis made on CoRoT red
giants, especially the mass and radius distributions \citep[Fig.
12 and 13 of][]{2010A&A...517A..22M}. Benefitting from the same
calibration of the asteroseismic mass and radius determination as
done in this work, we have investigated the mass-radius relation,
having in mind that without information of the evolutionary
status, there is no clear information \citep{2011MNRAS.tmp..559H}.

We note here that the mass distribution is almost uniform in the
RGB, contrary to clump stars (Fig.~\ref{MR}). The number of
high-mass stars in the RGB, above $1.8\,M_\odot$ as defined in
\cite{2011Natur.471..608B}, is about half of that in the clump,
consistent with their expected more rapid evolution. Similarly,
stars with masses below $1 \,M_\odot$ are significantly rarer (by
a factor of six to one) in the RGB by comparison with the clump.
If we assume that the scaling relations, valid along the whole
evolution from the main sequence to the giant class, remain valid
after the tip of the RGB, this comparative study proves that
low-mass stars present in the clump but absent in the branch have
lost a substantial fraction of their mass due to stellar winds
when ascending up to the tip of the RGB. Therefore, after the
helium flash, these stars show lower mass. The quantitative study
of this mass loss will require a careful unbiased analysis, out of
the scope of this paper. On the contrary, high-mass stars, even if
they lost mass too, can be observed as clump stars since they
spend a much longer time in the core-helium burning phase than
while ascending the RGB. Most of these stars belong to the
secondary red clump
\citep{1999MNRAS.308..818G,2011Natur.471..608B}. We note that the
secondary-clump stars present a larger spread in the mass-radius
distribution, certainly due to the fact that those stars that have
not undergone the helium flash present different interior
structures. On the other side, the low-mass stars of the clump
present a clear mass-radius relation: the helium flash shall have
made their core largely homologous.

\subsection{Mode amplitudes and number of modes}

We can also address the influence of the evolutionary status on
the energetic parameters of the red-giant oscillation spectrum.
When plotting the mean height of the excess power Gaussian
envelope as a function of the frequency $\numax$, we remark that
clump stars and RGB stars present very similar height for $\numax$
less than 40\,$\mu$Hz. However, above 40\,$\mu$Hz, the
oscillations in clump stars present systematically lower heights
than RGB stars (Fig.~\ref{height}). The contrast is more than a
factor of 2. On the other side, the number of modes, estimated
from the ratio $\dnuenv/\Dnu$ where $\dnuenv$ is the full width at
half-maximum of the total oscillation excess power envelope, is
similar below 40\,$\mu$Hz, but more than 30\,\% larger for clump
stars above this limit (Fig.~\ref{envel}). Despite the somewhat
arbitrary limit in $\numax$ we see here evidence of the secondary
clump \citep{1999MNRAS.308..818G}, which consists of He-core
burning stars massive enough to have ignited He in non-degenerate
conditions. These stars have, at $\numax > 40\,\mu$Hz, similar
radii as RGB stars, and larger masses. Hence, they have a smaller
$L/M$ ratio, so that the excitation of oscillation is expected to
be weaker. With a larger $\numax$, they also present a solar-like
oscillation spectrum with modes excited in a broader frequency
range.

We suggest that, in case the measurement of $\tgobs$ is made
difficult by a low signal-to-noise ratio, the oscillation
amplitude and the size of the Gaussian envelope with noticeable
amplitude may be used to help the determination of the
evolutionary status.

Finally, we note that the secondary-clump population is certainly
underestimated, owing to the fact that the low amplitudes make
their detection quite complex in CoRoT data. In fact, such stars
are observed with a low signal-to-noise ratio since their Fourier
spectra are dominated by a white-noise contribution.

\subsection{Different populations}

The number of detections of solar-like oscillation towards the
anticenter direction is lower than towards the center, due to a
lower red-giant density in the outer galactic regions and to
dimmer magnitudes \citep{2010A&A...517A..22M}. Hence, the number
of reliable measurements of $\tgobs$ is lower too, but with the
same proportion of 40\,\%. The sample is large enough to make sure
that the difference between the distributions is significant
(Fig.~\ref{scalingT}).

The anticenter field shows a significant deficit of red-clump
stars below $\Dnu=3.8\,\mu$Hz, e.g. with small mass. Another
similar deficit is observed at low frequency in the red-giant
branch; it should indicate populations with different ages. We
also note that secondary-clump stars of the anticenter direction
present slightly higher g-mode spacings; it should indicate
populations with different mass distributions. This illustrates
the interest to compare the different fields in order to analyze
different populations (Miglio et al., in preparation).

\section{Conclusion\label{conclusion}}

The clear identification of the p-mode oscillation pattern has
allowed us to identify in CoRoT observations the pattern of mixed
modes behaving as gravity modes in the core and pressure modes in
the envelope. We have verified that this pattern is very close to
the asymptotic expression of g modes. Benefitting from the
identification of $\ell=1$ mixed modes, we also have measured
$\ell=2$ mixed modes. The presence or absence of these mixed modes
will allow us to study the deep envelope surrounding the core.

We have verified that the mean large period separation of the
mixed modes gives a clear indication of the evolution of the star.
Assuming the validity of the asteroseismic scalings laws for the
stellar mass and radius, we have shown that the mass distribution
of the RGB is much more uniform than in the red clump.
Asteroseismology confirms that these red-clump  stars have
undergone a significant mass loss. Furthermore, red-giant low-mass
stars after the helium flash do present homologous interiors and a
clear mass-radius relation, contrary to stars in the secondary
clump. Longer time series recorded with \Kepler\ will allow us to
investigate this relation in more detail. Benefitting from the
fact that CoRoT provides observations in two fields, towards the
Galactic center and in the opposite direction, we have now a
performing indicator for making the population study more precise.
This will be done in future work.

These data confirm the power of red-giant asteroseismology: we
have access to the direct measure of the radiative central
regions. Even if observations only deliver a proxy of the large
period separation, comparison with modelling will undoubtedly be
very fruitful. In a next step, the dedicated analysis of the best
signal-to-noise ratio targets will allow us to sound in detail the
red-giant core. Modulation of the large period spacing, observed
in many targets, will give a precise view of the core layers.

\begin{appendix}

\section{Method\label{method}}

\begin{figure*}
\includegraphics[width=17.5cm]{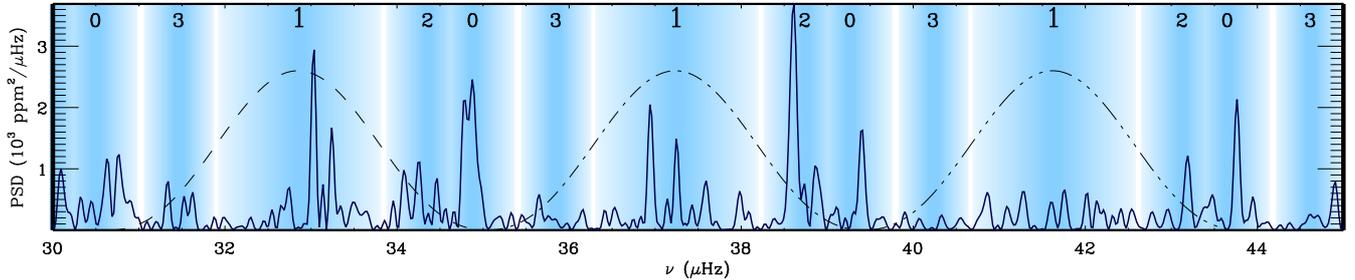}
\caption{Zoom on the bar code spectrum of the target CoRoT
100752538, with the large separation provided by the adjustment
derived from the red-giant oscillation universal pattern.
Different narrow filters centered on different pure $\ell=1$
pressure modes are indicated with different line styles. They
allow us to measure a local g-mode frequency spacing in each
filter.\label{filtre}}
\end{figure*}

Identifying $\ell=1$ mixed modes first requires to aim at them
precisely. This first step is achieved by the method presented in
\cite{2011A&A...525L...9M}, able to mitigate the major sources of
noise that perturb the measurement of the large separation $\Dnu$
and then to derive the complete identification of the p-mode
oscillation pattern. \emph{Complete identification} means that all
eigenfrequencies, their radial order and their degree, are
unambiguously identified, as for instance the expected frequencies
of the pure pressure $\ell=1$ modes:
\begin{equation}\label{ridge}
\nu_{n,\ell=1} = [n+1/2+\varepsilon(\Dnu) - d_{01}] \;\Dnu,
\end{equation}
with $\varepsilon(\Dnu)$ representing the surface term and
$d_{01}$ accounting for the small separation of $\ell=1$ pure p
modes.

\begin{figure}
\includegraphics[width=8.9cm]{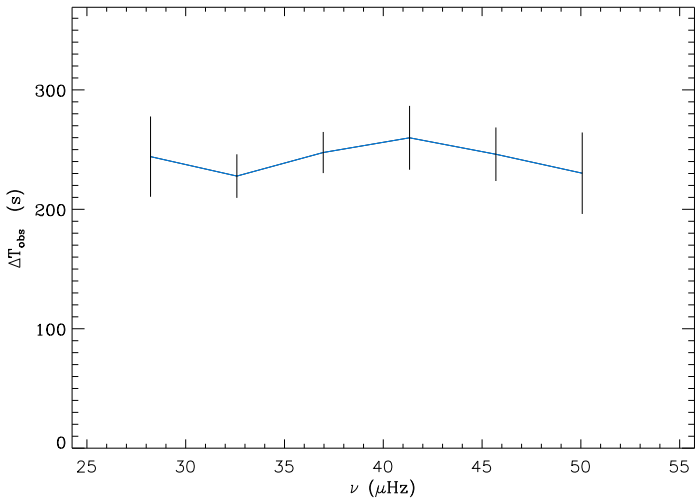}
\caption{Mean period spacings measured around the expected pure
$\ell=1$ pressure modes of the target CoRoT 100752538. Vertical
lines indicate the error bars on each individual measurement.
 \label{spacingok}}
\end{figure}

The frequency spacings of the mixed modes are then analyzed with
the envelope autocorrelation function (EACF) based on narrow
filters centered on the frequencies $\nu_{n,1}$ in the vicinity of
the frequency $\numax$ of maximum oscillation amplitude
\citep{2009A&A...508..877M}. We deliberately chose the EACF method
since it has proved to be efficient at very low signal-to-noise
ratio
\citep{2009A&A...506...33M,2010A&A...524A..47G,2011A&A...525A.131H},
thanks to a statistical test of reliability based on the null
hypothesis.

In order to only select mixed modes, the full width at
half-maximum of the filter is fixed to $\Dnu /2$
(Fig.~\ref{filtre}). From the differentiation of the relation
between period and frequency, the regular spacing of
Eq.~\ref{tassoul_un} in period translates into spacings varying
with frequency as:
\begin{equation}
\spacing =  \nu_{\ng,1}^{2}\ \tgun \simeq  \nu_{n,1}^{2}\ \tgun .
\label{spacing}
\end{equation}
For an individual measure centered on a given p mode, the
frequency spacings selected by the narrow filter can be considered
as uniform, since the ratio $\Delta \ng / \ng$ that represents the
relative variation of the gravity radial order within the filter
is less than about 1/25 (Table \ref{param}). Then, when comparing
the different measures around different pressure radial orders,
obtaining mean frequency spacings varying as $\nu_{n,\ell=1}^{2}$
validates the hypothesis of a Tassoul-like g-mode pattern
(Fig.~\ref{spacingok}).

We have checked that the method can operate with a filter narrow
enough to isolate the mixed modes. This is clearly a limit, since
the performance of the EACF varies directly with the width of the
filter \citep{2009A&A...508..877M}. We have also checked that such
a filter is able to derive the signature of $\ell=1$ mixed modes
in a frequency range as wide as possible, without significant
perturbation of possible $\ell=2$ mixed modes. As filters with a
narrower width focus too much on the region where, due to the
vicinity of the $\ell=1$ pure pressure mode, bumped mixed modes
present a narrower period spacing than that expected from
Eq.~\ref{tassoul_un}, we consider the $\Dnu/2$ width to be the
best compromise.

Since measurements can be verified in different frequency ranges
(Fig.~\ref{filtre} and \ref{spacingok}), we have chosen a
threshold value corresponding to the rejection of the H0
hypothesis at the 10\,\% level. For the characteristics of mixed
modes, different to the characteristics of the p modes considered
in \cite{2009A&A...508..877M}, this corresponds to a normalized
EACF of about 4.5 at $\numax$. In practice, stellar time series
with a low signal-to-noise ratio are excluded by this threshold
value. In case of reliable detection, error bars can be derived
following Eq.~A.8 of \cite{2009A&A...508..877M}.

The measurement of the g-mode frequency spacing $\spacing$ is not
only validated by a correlation signal larger than the threshold
level: we only selected results with $\spacing$ measured in at
least two frequency ranges and verifying Eq.~\ref{spacing} within
20\,\%.
This flexibility allows us to account for possible discrepancy to
the exact asymptotic relation (Eq.~\ref{tassoul_un}), as produced
by the avoided crossings resulting from coupling of the p mode in
the stellar envelope to the g modes in the core. It also accounts
for the possible modulation of the period due to a composition
gradient in the core \citep{2008MNRAS.386.1487M}.

Deriving an estimate $\tgobs$ of $\tgun$ from the spacings
$\spacing$ is then direct. Due to avoided crossings, $\tgobs$ is
close to but less than $\tgun$ \citep{2010A&ARv..18..471A,
2011Sci...332..205B,2011Natur.471..608B}. This is called mode
bumping and results from the fact that the mixed modes around
$\nu_{n,1}$ present necessarily smaller spacings than pure g modes
since the mixing of the g modes with one p mode gives one
supernumerary mixed mode per $\Dnu$ frequency interval, as shown
in \cite{2011Sci...332..205B}. The ratio $\tgun/\tgobs \simeq
1.15$ is derived from the examination, when possible, of the
g-mode spacing far from the expected p mode, assuming as in
\cite{2011Natur.471..608B} that this spacing is unperturbed by the
mode bumping and corresponds to the asymptotic g-mode spacing. The
ratio differs  from the value obtained with the Kepler data, since
the frequency resolutions and the analysis methods are different.
First, the frequency resolution is 2.5 times less fine for CoRoT
data; as a consequence, the influence of the mode bumping is more
smooth. Second, and more importantly, the envelope autocorrelation
method is able to derive a mean value of the spacing in a larger
frequency range than the method exposed in
\cite{2011Natur.471..608B} thanks to the $\Dnu/2$-broad filter
used to select the mixed modes. This helps to enhance the
contribution of non-bumped mixed modes.

When relaxing the condition expressed by Eq.~\ref{spacing}, the
method can be applied, with ultra-narrow filters centered on
individual mixed modes, to search for rotational splittings. This
search was unfortunately negative for CoRoT red-giant spectra.

\end{appendix}

\begin{acknowledgements}
This work was supported by the Centre National d'\'Etudes
Spatiales (CNES). It is based on observations with CoRoT. The
research has made use of the Exo-Dat database, operated at
LAM-OAMP, Marseille, France, on behalf of the CoRoT/Exoplanet
program. KB acknowledges financial support from CNES. SH
acknowledges financial support from the Netherlands Organisation
for Scientific Research (NWO). PB  received funding from the
European Community's 7th Framework Programme, ERC grant n°227224
(PROSPERITY).

\end{acknowledgements}

\bibliographystyle{aa} 
\bibliography{16825bib}

\end{document}